# Modeling of Chemical Vapor Infiltration Using Boundary Singularity Method

Authors: A. Povitsky (alex14@uakron.edu) and P. Mahoney (pwmahoney79@yahoo.com)

Department of Mechanical Engineering
The University of Akron, Akron, OH USA

**Abstract**

Boundary Singularity Method (BSM) was used to model Chemical Vapor Infiltration (CVI) in a fibrous preform. Straight, long fibers of varying cross-sectional geometry, representing fibers of a preform, were placed within a domain of a pre-determined size. The preparation of dense fiber-reinforced Silicon-Carbon (SiC) composites was considered as a representative of CVI methodology, where methyl-trichlorosilane (MTS) was used as both the silicon and carbon donor for the silicon carbide matrix. Concentrations of MTS were then set at the domain boundaries, and the domain was gradually infiltrated with MTS as time progressed. The concentration of MTS at the surface of the preform fibers was calculated using the adopted BSM. For quasi-equilibrium considered, the reaction rate at solid surface is equal to the diffusion rate towards the surface. The Robin or third type boundary condition, which is a linear combination of the values of a function and the values of its derivative on the boundary of the domain, are developed and implemented to BSM. From the fibers' surface concentrations obtained by BSM, deposition rates were calculated, and the geometry was updated to reflect the fiber growth during the time step, therefore, the fiber size growth and pore filling was modeled over time. The BSM analysis was verified by comparisons to a known analytical solution of concentric cylinders with a concentration set at the outer cylinder and a reaction at the inner. BSM solutions were also compared to experimental data as well as computational results obtained by a Level-Set Method (LSM). Obtained dynamics of pore size and location will help to evaluate quality of material manufactured by CVI. Porosity transients were obtained to show the relation between initial and current porosities as time progresses.

**Symbols**

| | |
|---|---|
| A | pre-exponential factor |
| C | concentration at boundaries |
| c(x,y) | local concentration |
| D | mass diffusivity |
| $E_a$ | activation energy |
| F | singularity strength |
| i | singularity index |
| j | collocation index |
| K | reaction rate constant |
| $M_{film}$ | molar mass of desired deposited material |
| n | normal direction |
| $n_x$ | unit normal x-component |
| $n_y$ | unit normal y-component |
| Pe | Peclet Number |
| $\rho_{film}$ | density of desired deposited material |
| R | ideal gas constant |
| r | radius |
| $r_i$ | inner radius |
| $r_o$ | outer radius |
| $r_{ij}$ | distance from singularity I to observation point j |
| t | time |
| T | temperature |
| u,v,w | Cartesian velocity components |
| x,y,z | Cartesian coordinates |

**Introduction**

Originally Chemical Vapor Infiltration (CVI) was developed as a way to densify porous graphite bodies [1]. Since its development CVI has been widely used in the development of composite materials to the point where more than half of all carbon/carbon composites are made by CVI [2]. The composites produced are often highly heat- and wear- resistant, and as a result they have been used in the aerospace [3] as a key enabling technology to reduce fuel consumption and emissions of gas turbine engines [4] and in manufacturing of heat exchangers and furnaces [5]. CVI is preferable to other methods of ceramic processing because it is a low stress, low temperature, easy to control process. Since it is a low stress and low temperature process it avoids many of the risks associated with high temperature processing.

At its basic form CVI involves introducing reactants into a porous preform. The reactants then form precursors which are deposited on the internal surfaces of the preform, causing the internal surfaces to grow and the solid portion of preform become thicker. As fibers grow and the preform becomes more infiltrated, some areas of the geometry will be closed off to reactant gas. This will lead to voids forming in the geometry and increased porosity when infiltration is completed. Accounting for these voids, which could be detrimental to quality of manufactured products, is imperative in modeling the CVI.

Modeling of CVI processes is important to predict infiltration, whether voids may form, and where. The goal of this paper is to introduce Boundary Singularity Method (BSM) to model CVI of representative nests of fibers representing a portion of an actual preform. Numerical experiments show that, in contrast with other boundary element methods, the BSM (also known as the Method of Fundamental Solutions (MFS)) requires relatively few boundary points and singularities to produce accurate results [6]. This will be done by solving the diffusion equation with surface chemical reaction, using the Green's functions for the Laplace operator to solve the concentration throughout the domain. From the concentrations at the fiber surfaces, the fiber growth rate will be calculated, and then the geometry will be updated to reflect the calculated fiber growth. BSM has not been used to model CVI in prior studies, to the best of the authors' knowledge.

Comparisons will be made to published experimental and computational solutions to verify the effectiveness of this methodology. There are many different CVI processes used in industry today which employ many different chemical reactions in order to deposit the desired end product on the preform surfaces [7]. Many of these reactions involve several steps before the desired material is deposited. Each reaction step and product would have different reaction rate, *k,* and diffusion coefficient, *D,* values. The simplified surface chemical reaction which will be discussed in this paper is a one-step isothermal irreversible reaction of transformation of gas methyl-trichlorosilane ($CH_3SiCl_3$, abbreviated as MTS) to solid silicon carbide (SiC) [8,9]. The fabrication of high temperature ceramic matrix materials is enabled by polymer-based precursors for infiltration of SiC preforms.

There are several methods used to transport reactants in a CVI process. These methods are classified by whether they use forced convection or diffusion, and whether thermal gradients are imposed [10]. The most widely used CVI is isobaric and isothermal for which transport occurs entirely by diffusion [7,11]. This type of CVI is modeled in the current study. For an isothermal CVI process, near-surface pores tend to close early in the process restricting gas flow to interior surfaces [12]. To counteract this effect machining and re-infiltration may be necessary to ensure the desired density. This results in longer production time and in more expensive end products.

CVI processes mentioned above are slow, taking on the order of weeks to infiltrate, where fiber growth is on the order of microns per hour. As a result of the time scales of the reaction time and the growth rates in the CVI processes being modeled, a quasi-steady approach was used in the current study. Most of current models are based on the pseudo-steady-state hypothesis [13]—the mass and momentum transfer is very rapid compared to changes in pore geometry. This quasi-steady approach involves solving the steady state Laplacian for concentration at every time step and then updating the geometries based on the calculated growth rates and time step increment.

Due to the complex geometries of the fibrous preforms which are infiltrated, simplified models of the considered processes have been developed. Many of these models are one-dimensional transient pore models. After these simplifications are made, many of these models can be solved analytically. An isothermal CVI reaction was modeled in frame of 1-D approach in [14]. A tapered pore was considered in [15]. The analytical solutions to the simplified pore models are limited to the infiltration of a singular round pore [7,16]. In terms of modeling CVI, BSM offers a clear advantage in the broad range of geometries which can be handled. Using BSM, an arbitrary fibers' cross section can be handled including irregular cross sections and nests of fibers in proximity.

Temperature gradient chemical vapor infiltration (TG-CVI) utilizes a thermal gradient to enhance diffusivity. The vapor initially diffuses to the hotter side of the preform and then to the cooler side as the hot side concentration increases. This process usually results in a more uniform densification because the preform densifies from the outlet to the inlet. Isothermal forced flow chemical vapor infiltration (F-CVI) forces the vapor through a uniformly heated preform by means of applied pressure gradient. The rate of deposition is increased by the forced infiltration of the reactant vapor. Both TG-CVI and F-CVI can be combined and the reactant vapor can be forced through a preform with a thermal gradient in order to speed up the infiltration process and ensure uniform densification. The TG-CVI and F-CVI will be modeled in the future by the extension of BSM introduced in the present study to convection-diffusion equation for F-CVI and by coupling Laplace equations for concentration and temperature for TG-CVI.

This paper is composed as follows. The second section contains governing equations, methodology used for their solution, and verification of this methodology by comparison to an analytical solution for diffusion between round surfaces and deposition at one of the surfaces. The third section discusses the obtained computational results as well as comparison to similar computational results calculated using the Level-Set Method (LSM) and a qualitative comparison to experimental results. Transient porosity and circumferential variation of deposition, which are difficult to obtain analytically, are computed by the proposed mathematical model in this section. The final section gives a conclusion reached as well as possible areas of further applications and developments of this study.

**2. Theoretical model and numerical methodology**

In CVI modeling it is necessary to solve the diffusion equation in order to obtain concentration of species in the computational domain as well as surface chemical reaction and growth rate resulting from deposited material. Distribution of species concentration could be described using the following concentration transport equation:

$$\frac{\partial c}{\partial t} + u\frac{\partial c}{\partial x} + v\frac{\partial c}{\partial y} + w\frac{\partial c}{\partial z} = D\left[\frac{\partial^2 c}{\partial x^2} + \frac{\partial^2 c}{\partial y^2} + \frac{\partial^2 c}{\partial z^2}\right], \quad (1)$$

where D is the diffusion coefficient.

As previously mentioned, the infiltration of the geometry is a very slow process. As a result, instead of modeling the actual transient process a quasi-steady approach was taken where the change of domain concentration over each individual time step is treated by a steady state analysis. The fiber size changes because of deposition are the geometry is updated after every time step. Since concentration update at each time step is obtained by steady state analysis, the term $\frac{\partial c}{\partial t}$ can be eliminated in above equation. Also, in this study geometries involving fibers in cross-section *(x, y)* will be considered. These geometries of fibers are assumed uniform along the third, z-axis, and as a result the $\frac{\partial c}{\partial z}$ terms can be eliminated in above equation. In the non-dimensional form and with above listed simplifications, the following equation can be written:

$$u\frac{\partial c}{\partial x} + v\frac{\partial c}{\partial y} = \frac{1}{Pe}\left[\frac{\partial^2 c}{\partial x^2} + \frac{\partial^2 c}{\partial y^2}\right], \quad (2)$$

where Pe is the Peclet number.

For considered CVI geometry the characteristic length is very small and the velocities are also small, that leads to very small Peclet number:

$$Pe = \frac{u \cdot L}{D} \ll 1 \quad (3)$$

In this case, equation (2) can further be reduced to:

$$0 = \frac{1}{Pe}\left[\frac{\partial^2 c}{\partial x^2} + \frac{\partial^2 c}{\partial y^2}\right]. \quad (4)$$

Equation (4) is a Laplace operator which can be solved using free-space Green's functions with boundary singularities. The free-space Green's function for the Laplace operator in two dimensions, in which we seek to solve the differential equation on an unbounded spatial region, is [17]

$$c^j = \frac{1}{2\pi}\sum_{i=1}^{n} F^i \ln(r^{(i,j)}), \quad (5)$$

where F represents the strength of the singularity, *i* represents the index (location) of singularity, *j* represents the index (location) of the observation point, *c* represents the concentration at the observation point, and $r^{(i,j)} > 0$ is the distance between the observation point *j* and location of the singularity, *i*.

To find unknown strength of singularity, F, by solving linear system of equations the number of singularities , *1≤i≤n*, and observation points, *1≤j≤n*, is the same and equal to *n*. The linear system based on equations (5) is given below:

$$\begin{bmatrix} c^{(1)} \\ c^{(2)} \\ c^{(3)} \\ \vdots \\ c^{(n)} \end{bmatrix} = \frac{1}{2\pi} \begin{bmatrix} \ln(r^{(1,1)}) & \ln(r^{(1,2)}) & \ln(r^{(1,3)}) & \cdots & \ln(r^{(1,n)}) \\ \ln(r^{(2,1)}) & \ln(r^{(2,2)}) & \ln(r^{(2,3)}) & \cdots & \ln(r^{(2,n)}) \\ \ln(r^{(3,1)}) & \ln(r^{(3,2)}) & \ln(r^{(3,3)}) & \cdots & \ln(r^{(3,n)}) \\ \vdots & \vdots & \vdots & \ddots & \vdots \\ \ln(r^{(n,1)}) & \ln(r^{(n,2)}) & \ln(r^{(n,3)}) & \cdots & \ln(r^{(n,n)}) \end{bmatrix} \begin{bmatrix} F^{(1)} \\ F^{(2)} \\ F^{(3)} \\ \vdots \\ F^{(n)} \end{bmatrix} \qquad (6)$$

To avoid high condition numbers of linear systems to be solved, singularities are located outside of the computational domain (submerged) as shown in Fig. 2. Submergence of singularities for Stokes equations was discussed in [18,19] and references therein. If the problem set-up includes only the Dirichlet (or first-type) boundary condition, the left-hand side of equation (6) is known and one should solve linear system with respect to unknown vector of strength of singularities, *F*. Boundary conditions of this type correspond to observation points at domain boundaries, where boundary concentrations are given.

The concentration at the fibers surface exposed to the CVI chemical reaction can be calculated using the following equation [20] assuming the equality of reaction rate at solid surface, $kc_{wall}$, to the diffusion rate towards the surface, $D\left(\frac{\partial c}{\partial n}\right)_{wall}$:

$$c_{wall} = \frac{D}{k}\left(\frac{\partial c}{\partial n}\right)_{wall}, \quad (7)$$

where k is the reaction rate which is calculated below using the Arrhenius Equation [21]:

$$k = A \exp\left(\frac{E_a}{RT}\right), \quad (8)$$

where A is the pre-exponential factor, $E_a$ is the activation energy, R is the ideal gas constant, and T is the reactor temperature. Since the CVI processes being modeled using BSM were isothermal, for a given reaction k is assumed to be a constant and calculated using (8). As can be seen from equation (8), reaction rate at different reactor temperatures will be different.

Equation (7) represents the Robin or third type boundary condition that is a linear combination of the values of a function and the values of its derivative on the boundary of the domain. The BSM is rarely used with this type of boundary condition. One example is BSM implementation of partial slip boundary conditions for Stokes equations [18]. To implement this boundary condition for equation (7), one need to take derivative of (5). For instance, derivative of (5) by *x* is expressed as follows:

$$\frac{\partial c}{\partial x} = \frac{1}{2\pi}\sum_{i=1}^{n} F^i \frac{\partial\left(\ln(r^{(i,x)})\right)}{\partial x} = \frac{1}{2\pi}\sum_{i=1}^{n} F^i \frac{1}{r^{(i,x)}} \frac{\partial r^{(i,x)}}{\partial x} = \frac{1}{2\pi}\sum_{i=1}^{n} F^i \frac{x - x^i}{r^{(i,x)2}}, \quad (9)$$

where $r^{(i,x)} = \sqrt{(x^i - x)^2 + (y^i - y)^2}$ is the distance between location of the i th singularity, $(x^i, y^i)$, and an arbitrary point within the computational domain, (x, y). Consequently, $\frac{\partial r^{(i,x)}}{\partial x} = \frac{x - x^i}{r^{(i,x)}}$.

The concentration derivative normal to the fibers' surface in equation (7) at observation points j is calculated by taking the directional derivative of (6) and substituting the coordinate of observation points, $x = x^j$, into equation (9). To calculate the directional derivative, first the derivative with respect to both x and y Cartesian coordinates is calculated:

$$\begin{bmatrix} \dfrac{\partial c^{(1)}}{\partial x} \\ \dfrac{\partial c^{(2)}}{\partial x} \\ \dfrac{\partial c^{(3)}}{\partial x} \\ \vdots \\ \dfrac{\partial c^{(n)}}{\partial x} \end{bmatrix} = \dfrac{1}{2\pi} \begin{bmatrix} \dfrac{r_x^{(1,1)}}{r^{(1,1)2}} & \dfrac{r_x^{(1,2)}}{r^{(1,2)2}} & \dfrac{r_x^{(1,3)}}{r^{(1,3)2}} & \cdots & \dfrac{r_x^{(1,n)}}{r^{(1,n)2}} \\ \dfrac{r_x^{(2,1)}}{r^{(2,1)2}} & \dfrac{r_x^{(2,2)}}{r^{(2,2)2}} & \dfrac{r_x^{(2,3)}}{r^{(2,3)2}} & \cdots & \dfrac{r_x^{(2,n)}}{r^{(2,n)2}} \\ \dfrac{r_x^{(3,1)}}{r^{(3,1)2}} & \dfrac{r_x^{(3,2)}}{r^{(3,2)2}} & \dfrac{r_x^{(3,3)}}{r^{(3,3)2}} & \cdots & \dfrac{r_x^{(3,n)}}{r^{(3,n)2}} \\ \vdots & \vdots & \vdots & \ddots & \vdots \\ \dfrac{r_x^{(n,1)}}{r^{(n,1)2}} & \dfrac{r_x^{(n,2)}}{r^{(n,2)2}} & \dfrac{r_x^{(n,3)}}{r^{(n,3)2}} & \cdots & \dfrac{r_x^{(n,n)}}{r^{(n,n)2}} \end{bmatrix} \begin{bmatrix} F^{(1)} \\ F^{(2)} \\ F^{(3)} \\ \vdots \\ F^{(n)} \end{bmatrix} \quad (10\ a)$$

$$\begin{bmatrix} \dfrac{\partial c^{(1)}}{\partial y} \\ \dfrac{\partial c^{(2)}}{\partial y} \\ \dfrac{\partial c^{(3)}}{\partial y} \\ \vdots \\ \dfrac{\partial c^{(n)}}{\partial y} \end{bmatrix} = \dfrac{1}{2\pi} \begin{bmatrix} \dfrac{r_y^{(1,1)}}{r^{(1,1)2}} & \dfrac{r_y^{(1,2)}}{r^{(1,2)2}} & \dfrac{r_y^{(1,3)}}{r^{(1,3)2}} & \cdots & \dfrac{r_y^{(1,n)}}{r^{(1,n)2}} \\ \dfrac{r_y^{(2,1)}}{r^{(2,1)2}} & \dfrac{r_y^{(2,2)}}{r^{(2,2)2}} & \dfrac{r_y^{(2,3)}}{r^{(2,3)2}} & \cdots & \dfrac{r_y^{(2,n)}}{r^{(2,n)2}} \\ \dfrac{r_y^{(3,1)}}{r^{(3,1)2}} & \dfrac{r_y^{(3,2)}}{r^{(3,2)2}} & \dfrac{r_y^{(3,3)}}{r^{(3,3)2}} & \cdots & \dfrac{r_y^{(3,n)}}{r^{(3,n)2}} \\ \vdots & \vdots & \vdots & \ddots & \vdots \\ \dfrac{r_y^{(n,1)}}{r^{(n,1)2}} & \dfrac{r_y^{(n,2)}}{r^{(n,2)2}} & \dfrac{r_y^{(n,3)}}{r^{(n,3)2}} & \cdots & \dfrac{r_y^{(n,n)}}{r^{(n,n)2}} \end{bmatrix} \begin{bmatrix} F^{(1)} \\ F^{(2)} \\ F^{(3)} \\ \vdots \\ F^{(n)} \end{bmatrix} \quad (10\ b)$$

where $r^x = x^j - x^i$ and $r^y = y^j - y^i$ are the directional distances between the observation point $j$ and singularity point $i$ in the x and y directions, respectively.

To complete the directional derivative the dot product of (10a-b) is taken with the unit normal to surface vector $[n_x, n_y]$:

$$\frac{\partial c}{\partial n} = \left[\frac{\partial c}{\partial x}\ \frac{\partial c}{\partial y}\right] \cdot [n_x\ n_y] \quad (11)$$

By rearranging equation (6) and substituting in equation (11) for an observation point, $j$, the following equation at deposition surface is obtained:

$$0 = \frac{1}{2\pi} \begin{bmatrix} \ln(r^{(1,j)}) - \frac{D}{k}\left(\frac{r_x^{(1,j)}}{r^{(1,j)2}}n_x + \frac{r_y^{(1,j)}}{r^{(1,j)2}}n_y\right) \\ \ln(r^{(2,j)}) - \frac{D}{k}\left(\frac{r_x^{(2,j)}}{r^{(2,j)2}}n_x + \frac{r_y^{(2,j)}}{r^{(2,j)2}}n_y\right) \\ \ln(r^{(3,j)}) - \frac{D}{k}\left(\frac{r_x^{(3,j)}}{r^{(3,j)2}}n_x + \frac{r_y^{(3,j)}}{r^{(3,j)2}}n_y\right) \\ \vdots \\ \ln(r^{(n,j)}) - \frac{D}{k}\left(\frac{r_x^{(n,j)}}{r^{(n,j)2}}n_x + \frac{r_y^{(n,j)}}{r^{(n,j)2}}n_y\right) \end{bmatrix}^T \begin{bmatrix} F^{(1)} \\ F^{(2)} \\ F^{(3)} \\ \vdots \\ F^{(n)} \end{bmatrix} \quad (12)$$

Boundary conditions (12) are used for observation points located at fiber surfaces at which deposition occurs. For each observation point located at a boundary where deposition occurs, (12) is substituted into (6) in place of the original row, which corresponds to a given concentration.

To construct the domain, fibers were placed randomly in the domain one at a time until the desired porosity was achieved. It should be noted that even though the initial porosity of the sample is the same the actual geometry of the domain could be different due to the random way the fibers are being placed. The resulting linear system with respect to the strengths of the singularities, F, was then solved using the backslash operator in MATLAB [22].

To verify that the BSM methodology is solving the Laplace operator correctly for the given boundary conditions, an analytical solution for concentration between two concentric cylinders was developed. A diagram of this test case is shown in Figure 1. In 1-D cylindrical coordinates, the Laplace's equation is written as follows

$$0 = \frac{1}{r}\frac{\partial}{\partial r}\left(r\frac{\partial c}{\partial r}\right) \quad (13)$$

In this particular case the outer cylinder surface had a known constant concentration of $C$ and the inner cylinder, representing an isolated fiber, was a reaction boundary described by (7) (see Figure 1). The analytical solution of above equation with these boundary conditions is given by:

$$c(r) = \frac{C}{\frac{D}{k}\frac{1}{r_i} + \ln\left(\frac{r_o}{r_i}\right)}\ln(r) + \frac{C}{\frac{D}{k}\frac{1}{r_i} + \ln\left(\frac{r_o}{r_i}\right)}\left[\frac{D}{k}\frac{1}{r_i} - \ln(r_i)\right] \quad (14),$$

where $r_i$ and $r_o$ represent the inner and outer radius of cylindrical surfaces, respectively.

The set-up of the BSM solution is shown in Figure 2. The red markers show the location of the singularities, and the blue markers represent observation points where the boundary conditions are set-up. The observation points are placed on the inner and outer cylinder while the singularities are placed inside the cylinders at a submergence depth of 0.1 r, where r is the radius of the inner or outer cylinder, in correspondence. The number of singular points and collocation points is the same. Each surface has 40 observation points. The analytical solution (14) and the BSM solution calculated using the free space Green's function for the Laplace operator with corresponding boundary conditions (6, 12) are shown in Figure 3.

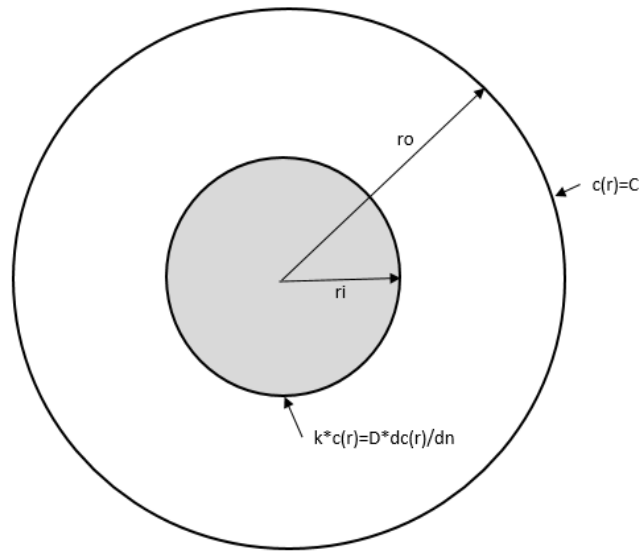

**Figure 1:** The set-up of diffusion between two cylinders. Concentration C at the outer cylinder surface is given. The inner cylinder surface has a reaction boundary at equilibrium.

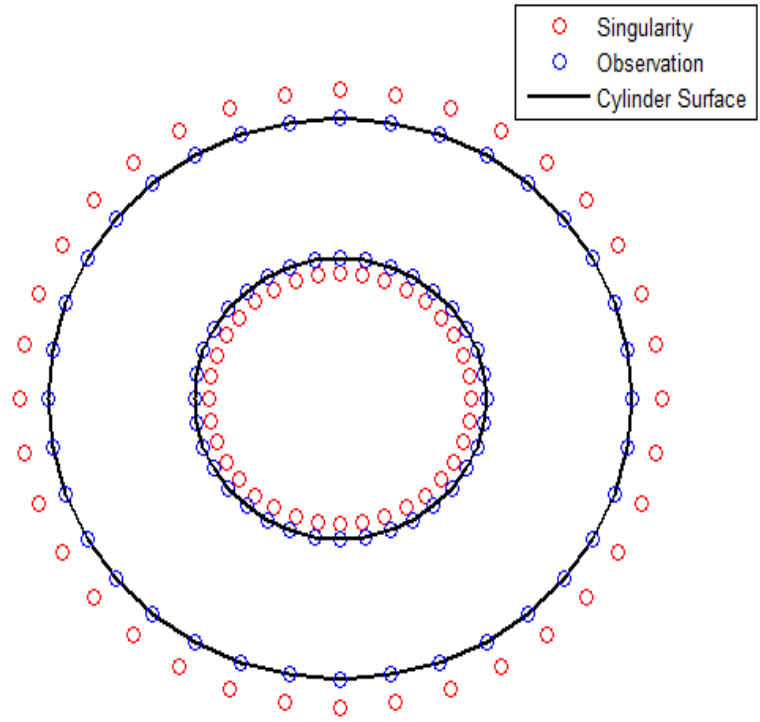

**Figure 2:** The BSM set-up for diffusion between two cylinders: red markers are submerged singularities while blue markers are observation points

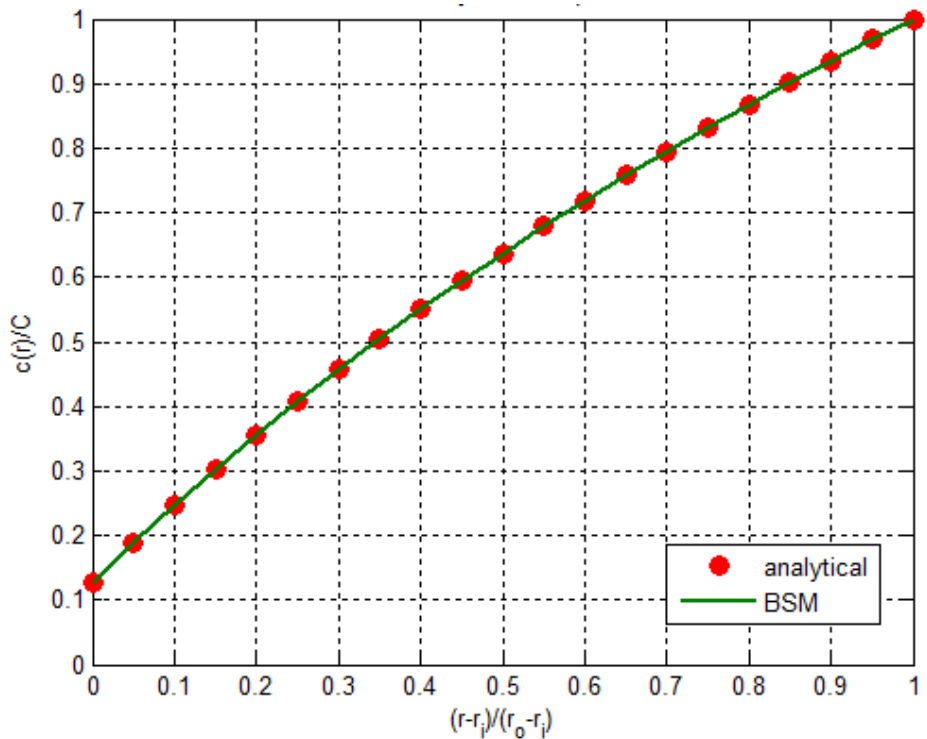

**Figure 3:** Comparison of the concentration c(r) obtained by BSM with the analytical solution (14)

After BSM solution for concentration for given geometry of fibers with deposited material (Figure 4), the concentration of the reactant at the wall can then be used to calculate the infiltration or growth rate [8]. Having calculated the concentration at the wall using (7), the growth rate of the fiber can then be calculated as follows by multiplying the concentration at the wall, the reaction rate calculated using the Arrhenius equation (8), and the ratio of the molar mass ($M_{reactant}$) and density of the reactant gas ($\rho_{reactant}$):

$$\delta = \frac{M_{reactant}}{\rho_{reactant}} \cdot k \cdot c_{wall} \quad (15)$$

Void tracking is an important part of modeling CVI. Figure 4 shows a depiction of a void. The area outlined in red is not accessible to reactant gas. These fibers' surfaces will not continue to grow during the rest of the simulation. To determine the location of these voids, the computational domain was meshed. The concentration boundaries were C>0 were then set to 1 and any mesh point within a fiber was assumed to be a C=0 boundary condition. The Laplace operator (4) for concentration was then solved using a finite-difference scheme. After the solution has converged any mesh point with C>0 has access to reactant and is not in a void while any mesh point with C=0 is either part of a fiber or in an area completely surrounded by fibers and has no access to reactant. Solving (4) using a finite difference scheme is done entirely for void tracking purposes and is not used to calculate concentrations or growth rates within the computational domain. This is a similar method to what was developed in [23].

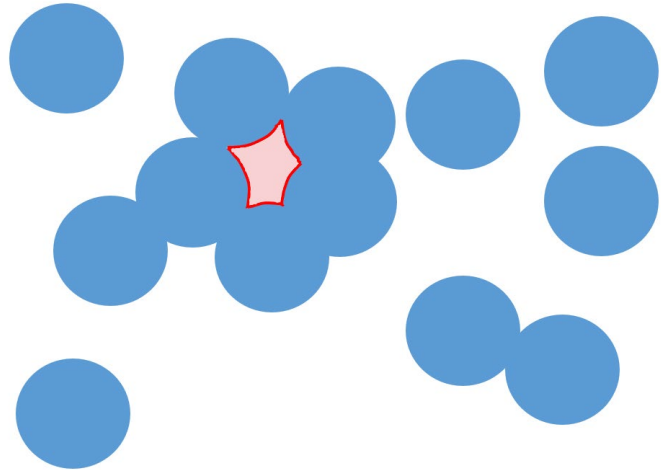

**Figure 4:** Example of void (red) in fibers' geometry with deposited material (blue).

Once the growth rate is obtained using (15), the singularity and observation points are moved in the normal direction to the local surface and the BSM solution is repeated at next time step. The process continues until either the sample has been fully infiltrated (porosity is no longer decreasing) or the specified end time is reached.

Porosity of the samples was calculated by generating 10,000 random points within the computational domain.  Whether the generated point was within a fiber or not was determined using the fiber center and observation points.  The number of total points in fibers divided by the total number of points is the specimen's porosity.  To determine the accuracy of this porosity measurement a sample with a known 80% porosity was measured 30 times.  The average of the measurements was 79.93% that was close to 80%. The porosity minimum and maximum were 79.23% and 80.73%, respectively.  The 3-σ range was 78.73%-81.27%.

## 3. Computational results and discussion

To further verify BSM as a way to model chemical vapor infiltration, comparisons were made to both experimental and computational results obtained by using LSM, and analytical solutions to simplified geometries.

In general, "real time" simulations of CVI experiments with the same number of fibers are not feasible at this time.  This is mainly due to computational limitations and the complex asymmetrical nature of the fiber preforms used.  To model a woven fiber preform using BSM millions of singularities would be required.  The resulting matrix which would need to be solved to get the strength of these singularities would as a result be on the order of $10^{12}$ entries.  This would need to be done multiple times as time progresses throughout the infiltration process.   As a result, the set-ups of fibers considered using computational techniques are much simplified versions of the experiments actually run such as unidirectional fiber-reinforced substrates [24].  A 3-D woven preform would need to be reduced to the hundreds of 2-D long cylinders randomly located in a computational domain representing a small section of the actual geometry. The computational results were generated by simulating a small section of a woven geometry used in a CVI experiment. The initial porosity of this sample was 38.81% and the final porosity after infiltration using BSM was 13.75%.   A qualitative comparison of computational results by using BSM (Figure 5a) to experimental results [25] (Figure 5b) can be found in Figure 5.  The BSM results show good qualitative co-incidence with experimental results.

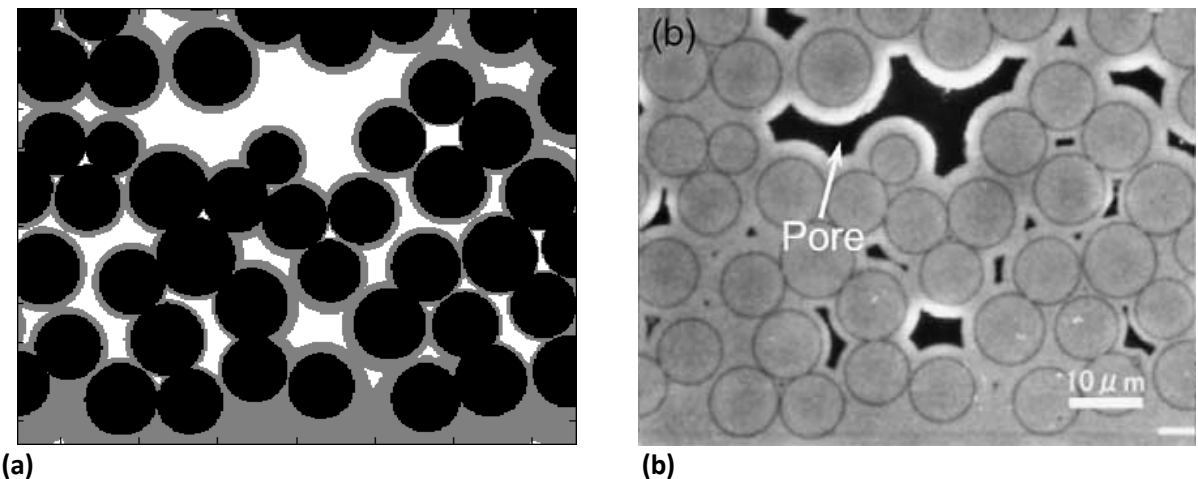

(a)         (b)

**Figure 5:**  Comparison of infiltrated geometry (a) computed by BSM and (b) obtained experimentally [25].

Next, the BSM results were compared to computational results using LSM, see [23] and references therein. The LSM [23] based on solution of Hamilton–Jacobi level set equation [26] is different from BSM in that it assumes a constant front speed/growth rate as a result of the transport of MTS from the surroundings being rapid relative to the growth rate. BSM has the advantage, which will be shown in the next section, of capturing both circumferential concentration and growth rate gradients, which will undoubtedly occur as a result of the geometry becoming more and more infiltrated.

The first geometry considered was 100 circular fibers arranged in an orthogonal 1 x 1 square, see Figure 6. All four boundaries were of a known concentration, and the initial fiber radius was 0.04. The time scale based on constant reaction rate, size of domain and fiber radius are taken from [23]. The exact initial porosity was 49.73% and the calculated initial porosity, using the methodology described in the past section, was 49.83%. The initial and the final infiltrated geometry are shown in Figure 6, and a transient of the infiltration process (porosity with respect to time) is shown in Figure 7. Using BSM, the final porosity was calculated to be 16.75%. The final porosity calculated using the LSM was 21.49% which compared well with the 21.46% by analytical solution [23].

This analytical solution was generated based on the assumption that all fibers grow at the same uniform rate. Using this assumption at final time moment all obstructions would be able to fit in a square box with its side equal to 2 r. The porosity of this box can then be calculated as the difference between the square and the cylinder, $Porosity = \frac{(2r)^2 - (\pi r^2)}{(2r)^2} = 1 - \frac{\pi}{4}$.

The difference in the final porosity calculation between BSM and LSM are in part due to different end criteria. The BSM method infiltration simulation was concluded when porosity no longer decreases (see Figure 7), while the LSM simulation concluded when all of the "outside" fibers had intersected and reactant could no longer reach fibers in the center of the domain.

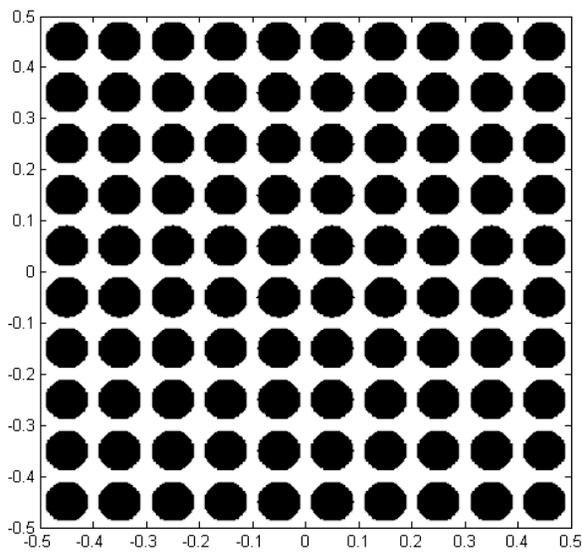

a.

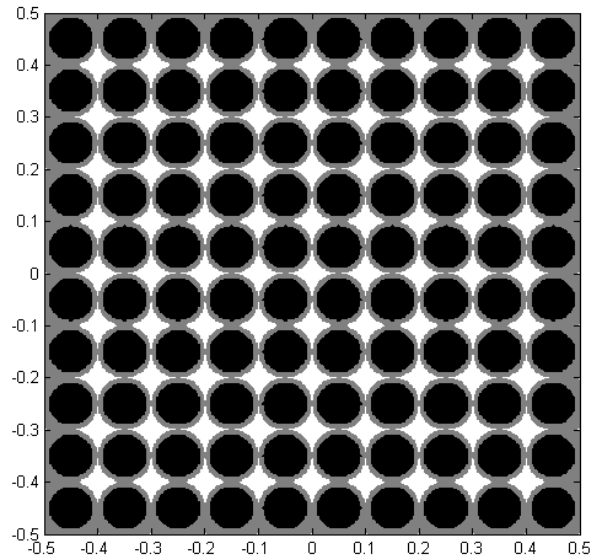

b.

**Figure 6:** Infiltration of fibers' set using BSM (a) initial geometry of fibers and b) final geometry after infiltration

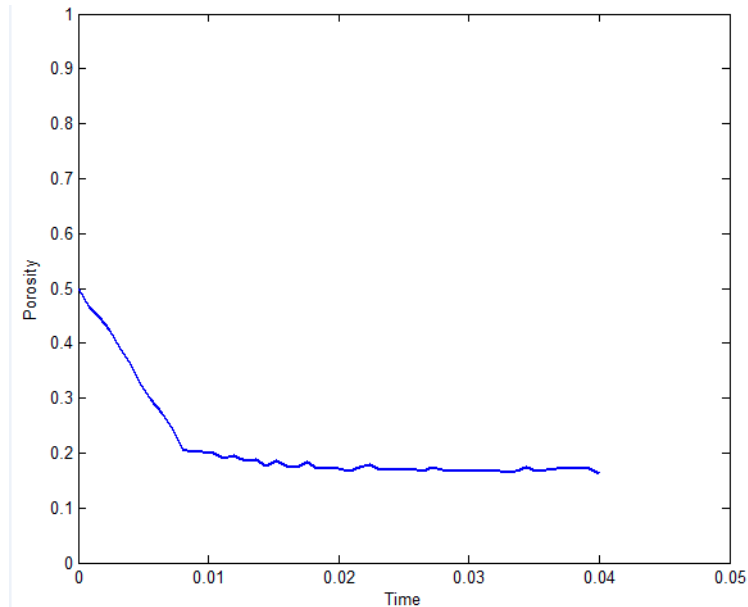

**Figure 7:** Porosity transient of aligned fibers' set

Infiltration simulations using offset fibers were also performed using BSM for offset circles, ellipses with a 1.5 aspect ratio, and squares. Results of these simulations are shown in Figure 8(a)-(c), in correspondence, where initial preform and final (completely infiltrated) geometries are shown. All of these geometries had the same initial porosity and involved 25 offset fibers. The offset circle sample was also simulated using the LSM in [23]. The final porosity using BSM was 5.42% which compared reasonably well to the LSM solution, 11.24%, and the analytical solution, 9.31%. It should be noted that the analytical and LSM solutions assume a uniform growth rate around each fiber, and some of the differences in final porosity are a result of different simulation conclusion criteria discussed above.

The transient infiltration process for offset fibers using BSM was modeled using the above-listed different cross-sectional shape fibers starting with the same initial porosity of 55%. Initially the infiltration process seems to be very similar for all of these geometries till porosity of approximately 22%. When fibers begin to intersect and voids begin to form, the infiltration process begins to vary. The final porosities offset circles was given above as 5.42%, while the final porosity of the elliptical cross section was 9.55%, and the rectangular cross section was 7.68%. Although the initial porosities were the same there is a substantial difference in the final porosities. This is because different geometries will create different size and shape voids.

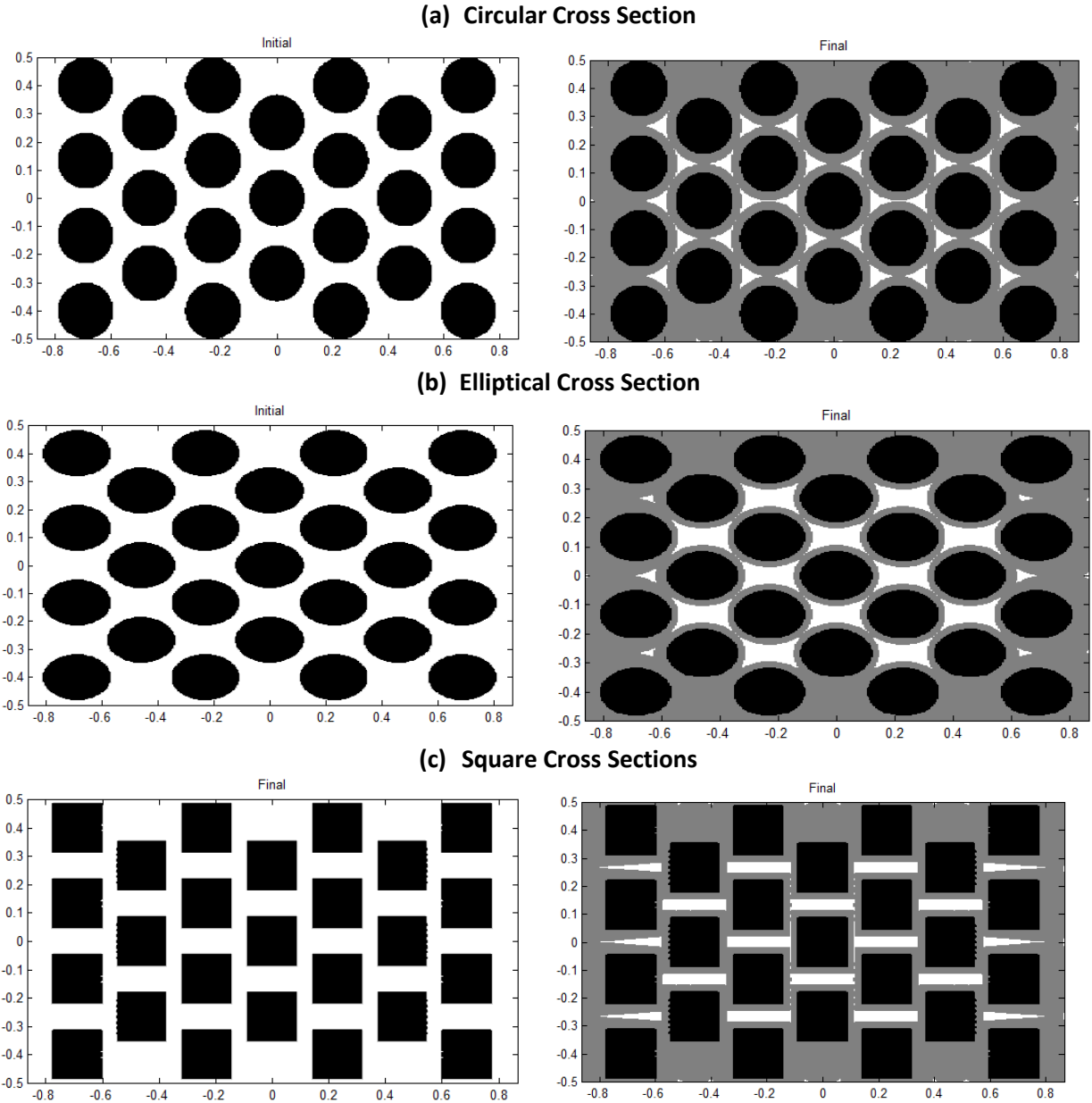

**Figure 8:** Infiltration of offset set-ups of fibers (a) circular, (b) elliptical, and (c) square fibers using BSM (right-initial geometry of fibers, left-final infiltrated and deposited geometry)

Further comparisons were also done to geometries of randomly generated circular fibers at different known porosities. These geometries were generated by randomly placing fibers in the domain one at a time until the desired porosity was reached. Geometries were generated with 80%, 70%, 60% and 50% initial porosity. A comparison of the final porosities between BSM and LSM [23] computational methods are shown in Table 1. Plots of the initial and final geometry of each sample infiltrated using BSM are shown in Figure 9, and the transients of the infiltration process are shown in Figure 10.

**Table 1:** Comparison of final porosity as a function of initial porosity of completely infiltrated geometry using BSM and LSM.

| Initial Porosity | 80% | 70% | 60% | 50% |
|---|---|---|---|---|
| Final Porosity (BSM) | 26.95% | 29.21% | 29.47% | 29.58% |
| Final Porosity (LSM) | 25.29% | 26.88% | 27.20% | 27.10% |

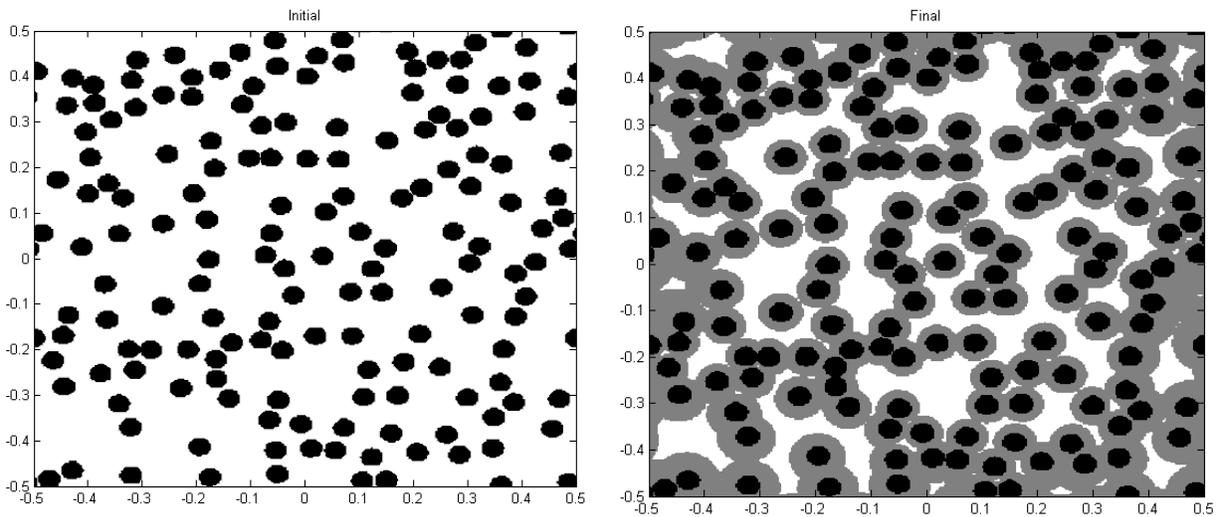

(a)

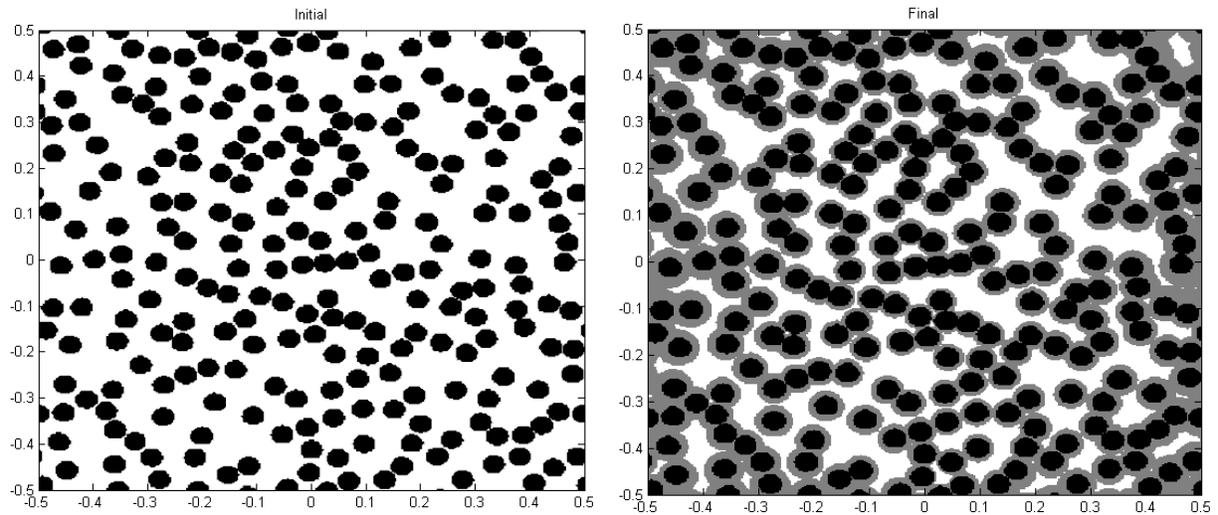

(b)

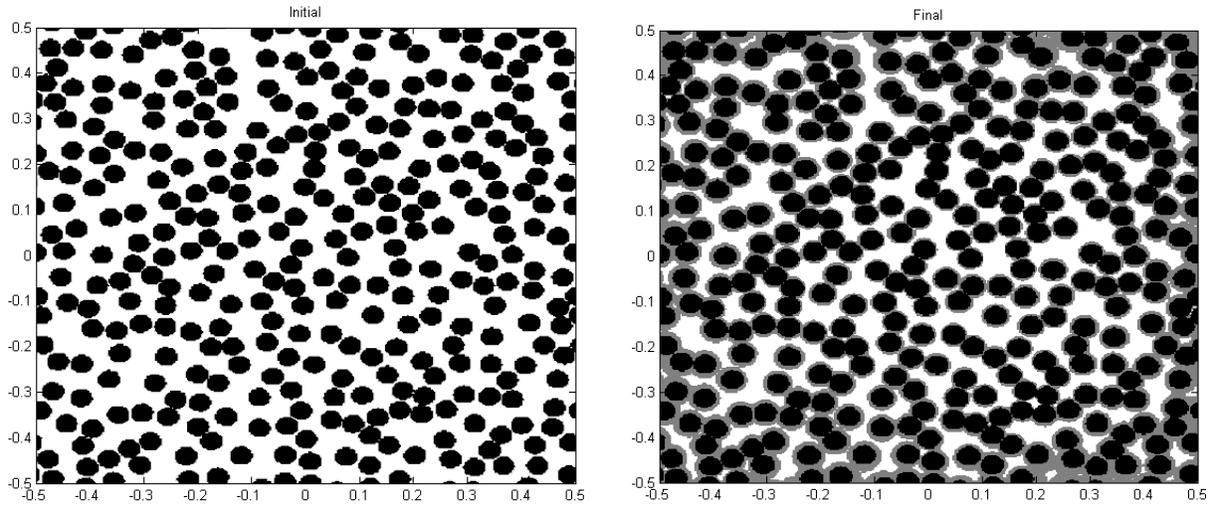

**(c)**

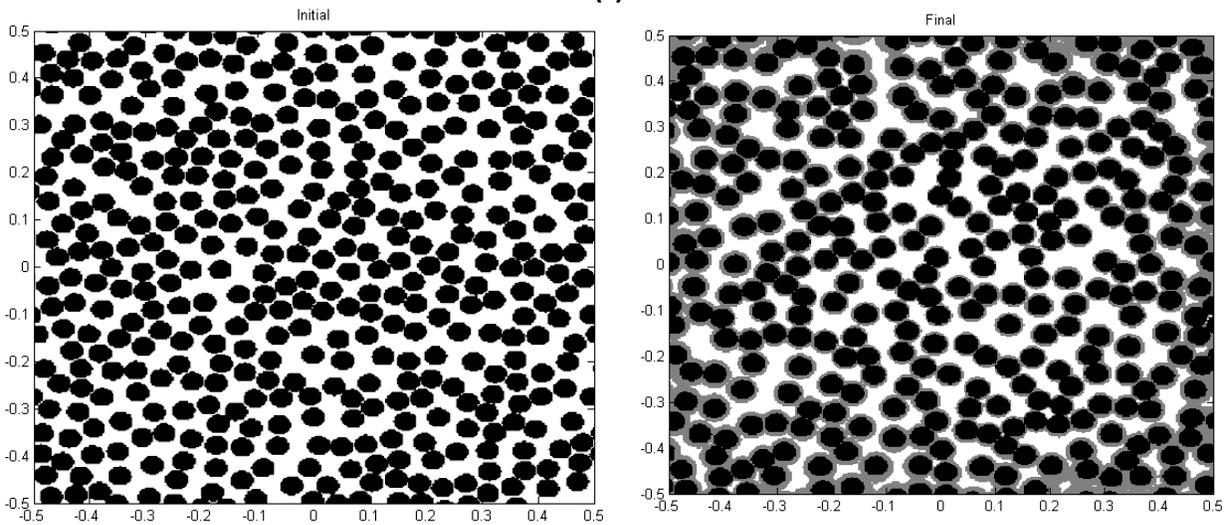

**(d)**

**Figure 9:** Infiltration of randomly placed fibers using BSM (right-initial preset geometry of fibers, left-final geometry after deposition with initial porosity of: a) 80%, b) 70%, c) 60%, and d) 50%.

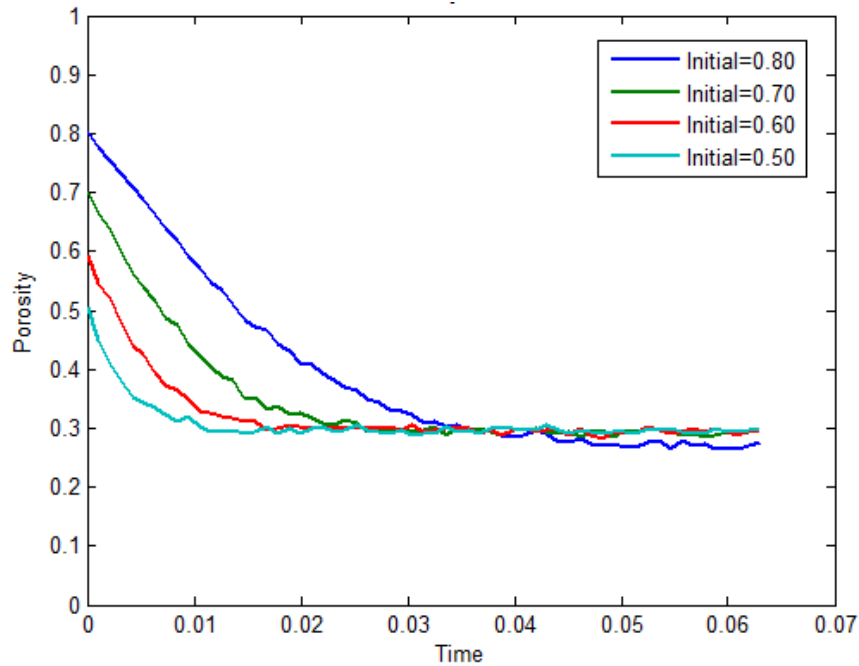

**Figure 10**: Porosity transient for randomly placed circular fibers with varying initial porosities

As previously mentioned the LSM approach presented in [23] assumes a constant front speed. It was assumed in [23] that the transport of MTS from the surroundings being rapid relative to the deposition rate and growth continuing at a constant rate as long as there is a continuous path to the precursor source. When the porosity is not too low this assumption is a valid estimate, however, as infiltration progresses variations of deposition rate along the circumference of fiber and throughout the domain can occur. The advantage of using BSM is that it does not require such an assumption.

As infiltration progresses and porosity decreases, reactant may become less and less available and as a result growth rates may vary from fiber to fiber and even circumferentially around a given fiber. Concentration at the boundaries where precursor diffuses into the domain can be much higher than that near the center. To show this, representative fibers were chosen from the 80% initial porosity sample, see Figure 11.

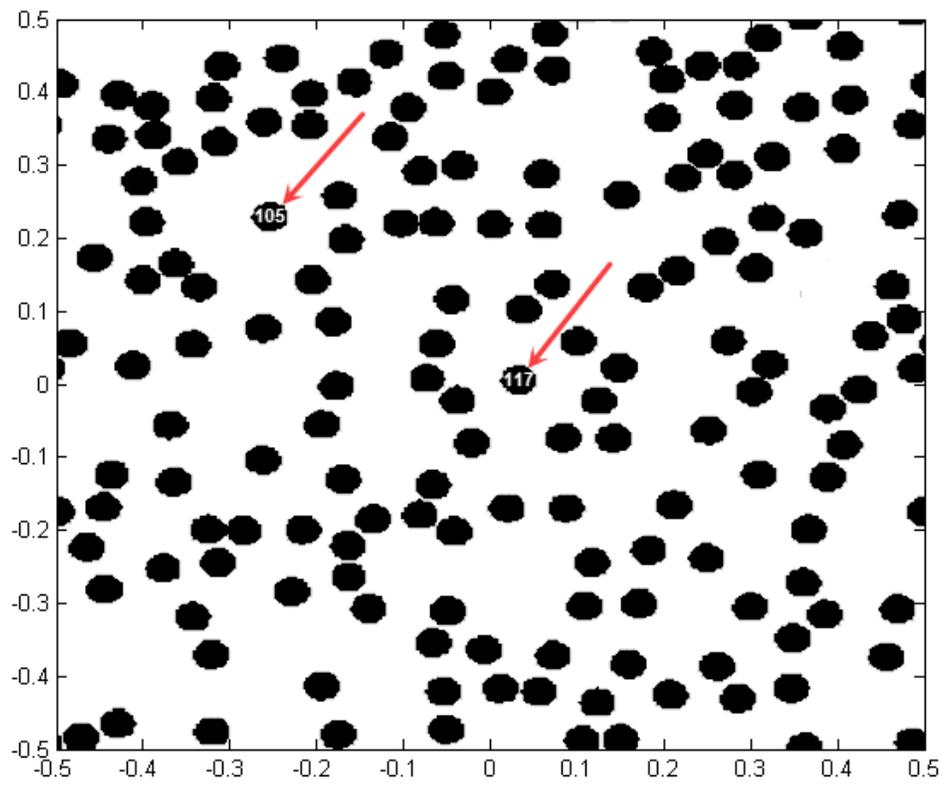

(a)

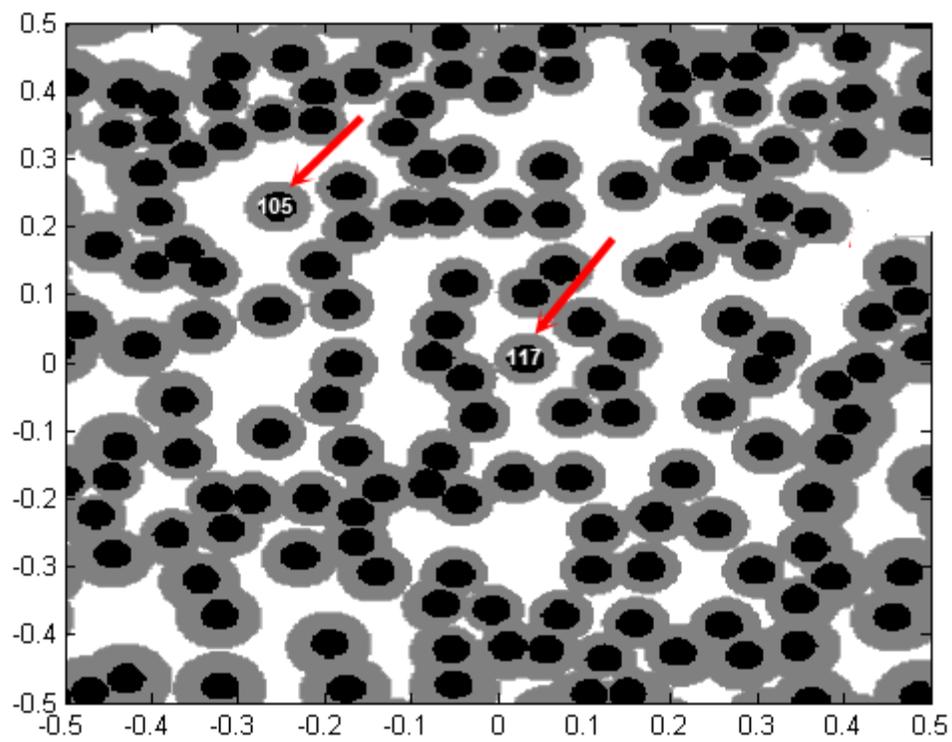

(b)

**Figure 11:** Location of fibers #105 and #117 within the computational domain at t=0 (a) and at the 80% initial porosity preform at t=0.0346 (b).

A transient plot of mean concentration at the fiber surface of #105 and #117 is shown in Figure 12. Initially the concentration at the domain and at its boundaries was equal to 1; therefore, the mean concentration about both fibers initially is equal to unity. At early time moment, the concentration throughout the domain is nearly uniform and the growth around all fibers is nearly uniform. As time progresses and the domain infiltrates, the concentration about these selected fibers begins to differ from that of the boundaries and that of each other. This will result in different fiber growth rates throughout the domain. As time progresses, the mean concentration about these constructions continues to drop and the concentration throughout the domain becomes less and less uniform. At approximately t=0.034 the mean concentration around both fibers drops to 0. At this point both fibers #105 and #117 no longer have access to precursor as a result of being entirely cut off from precursor diffusing in from the boundaries due to infiltration as shown in Figure 11(b). This is in contrast to the assumptions made in the LSM solutions which would result in a straight horizontal line near 1 until the fiber can no longer be reached by reactant diffusing from the boundaries.

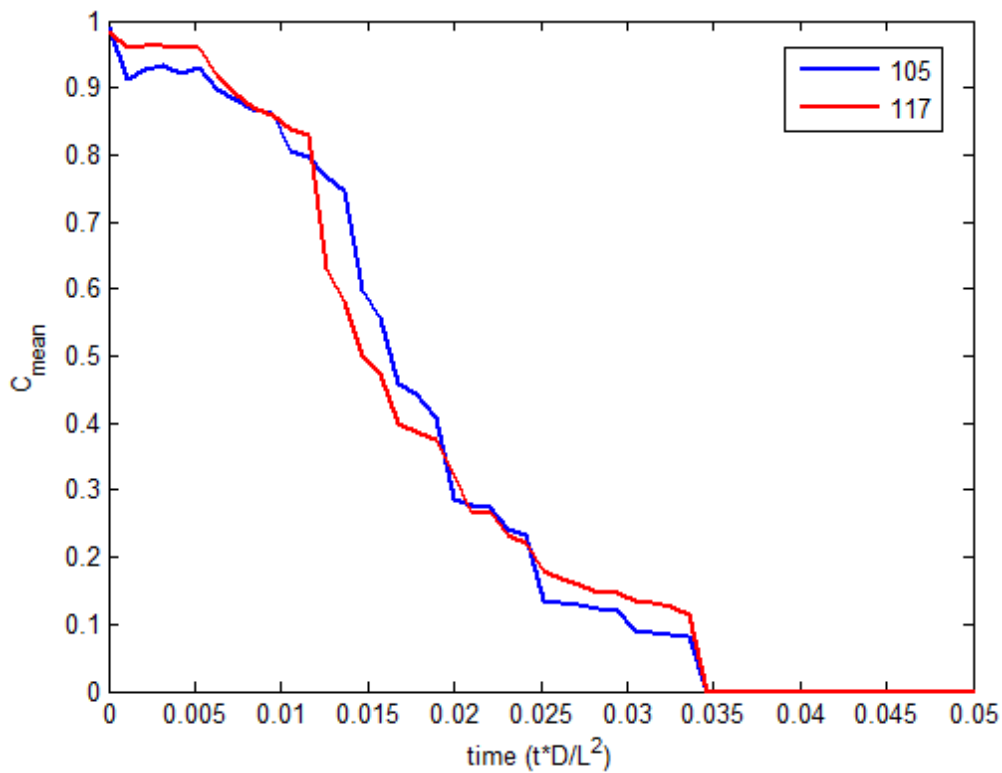

**Figure 12:** Mean concentration about selected fibers

Circumferential concentration also varies around fibers. This variation will lead to fibers growing from circular to varying degrees of elliptical cross-section. Figure 13 shows the 3-σ concentration variation about fibers #105 and #117. The standard deviation is calculated as:

$$\sigma = \sqrt{\frac{\sum_{i=0}^{n}|C - \bar{C}|^2}{n}} \quad (16)$$

Initially there is little to no variation circumferentially about these two fibers. As time and infiltration progress, circumferential variation also increases. At its most the 3-σ variation around fiber #105 is roughly 35% of the mean concentration about the fiber. Note that the value of σ is zero within LSM approach.

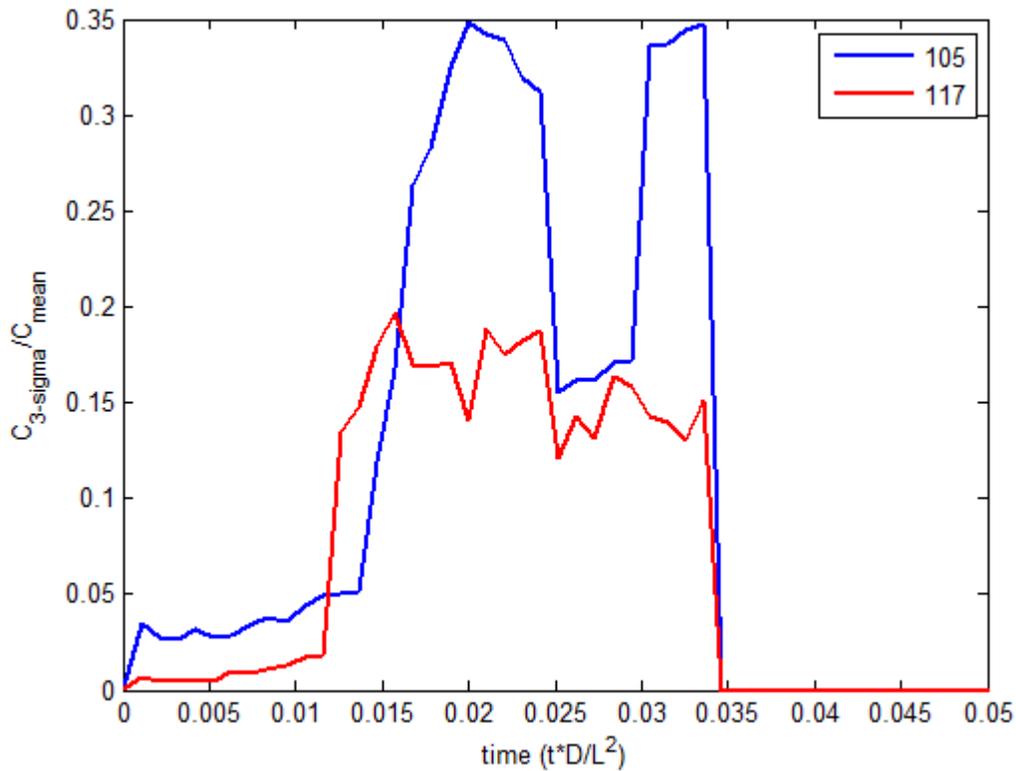

**Figure 13:** Circumferential variation of surface concentration about selected fibers

**Conclusions**

The BSM method was developed for Chemical Vapor Infiltration (CVI) with surface chemical reaction. BSM has not been used to solve CVI problems as of yet. The concentrations of reagent within the preform were obtained using the free-space Greens' functions for the Laplace operator describing diffusion with boundary conditions corresponding to the CVI reaction at fibers' surfaces. For quasi-equilibrium at fibers' surface considered, the reaction rate at solid surface is equal to the diffusion rate towards the surface. To account for this condition, the Robin or third type boundary condition at fibers' surface is developed and successfully incorporated into BSM. These surface concentrations were than used to calculate fiber growth rate due to deposition. Comparisons to analytical solution for simplified geometry and to the experimental snapshot of fibers' preform with deposition were the basis of evaluation of accuracy of BSM results.

The structured, offset and random placements of fibers with a given porosity were used as initial geometry conditions for CVI. CVI simulations using BSM were carried out for several preforms with a range of shapes and initial porosities. Computations by BSM in terms of porosity were compared to results obtained by the LSM in terms of porosity and patterns of deposition of infiltrated material. The difference in computational results coming from constant reaction rate assumption of LSM was discussed.

Dynamics of pore size and location were computed using the proposed BSM approach to evaluate quality of material obtained by CVI. Porosity transients were obtained for the range of initial porosities to show the relation between initial and final porosities and needed time to reach the final porosity. The circumferential deposition at fiber surface was obtained at different temporal stages of CVI to identify situations of non-uniform deposition depending on location of fibers relative to the domain boundaries and fibers' proximity to neighboring fibers.

The Boundary Singularity Method appears to be effective in modeling CVI with simplified one-step chemical surface reactions. The method accounts for circumferential variation of deposition rate along fibers' surface, depletion of chemical at the part of fibers' surface, merging of fibers with formation of voids caused by deposition and non-uniform distribution of reactant throughout the domain. Extension of the proposed approach to advection-diffusion equation will be considered in the future research.